\documentclass[a4paper]{article}

\usepackage[T1]{fontenc}
\usepackage[utf8x]{inputenc}
\usepackage[english]{babel}

\usepackage{cite}
\usepackage[super,sort&compress,comma]{natbib} 
\usepackage{amssymb,amsbsy}
\usepackage[intlimits,sumlimits]{amsmath}

\usepackage{droidsans}
\usepackage[colorlinks=true, allcolors=blue]{hyperref}

\newcommand{\email}[1]{\href{mailto:#1}{\tt{\nolinkurl{#1}}}}
\newcommand{\orcid}[1]{ORCID: \href{https://orcid.org/#1}{\tt{\nolinkurl{#1}}}}

\usepackage[sfdefault,lf]{carlito}
\usepackage[parfill]{parskip}

\usepackage{fancyhdr}
\usepackage{authblk}
\usepackage[a4paper,top=3cm,bottom=2cm,left=3cm,right=3cm,marginparwidth=1.75cm]{geometry}

\usepackage{amsmath}
\usepackage{graphicx}
\usepackage{booktabs}

\title{Phase Behavior of the Patchy Colloids Confined in the Patchy Porous Media}
\author[1,*]{Yurij V. Kalyuzhnyi}
\author[1,2]{Taras Patsahan}
\author[1]{Myroslav Holovko}
\author[3]{Peter T. Cummings}
\affil[1]{Institute for Condensed Matter Physics of the National Academy of Sciences of Ukraine, 1~Svientsitskii Street, UA-79011 Lviv, Ukraine}
\affil[2]{Lviv Polytechnic National University, 12~S.~Bandera Street, UA-79013 Lviv, Ukraine.}
\affil[3]{School of Engineering and Physical Sciences, Heriot-Watt University, Edinburgh EH14 4AS, United Kingdom}
\affil[*]{Corresponding author: \email{yukal@icmp.lviv.ua}}
\date{}

\begin{document}
	\maketitle

\begin{abstract}
A simple model for functionalized disordered porous media is proposed and the effects of confinement on self-association, percolation and phase behavior of a fluid of patchy particles are studied. The media is formed by a randomly distributed hard-sphere 
obstacles fixed in space and decorated by a certain number of off-center square-well 
sites. The properties of the fluid of patchy particles, represented by the fluid of hard
spheres each bearing a set of the off-center square-well sites, are studied using
an appropriate combination of the scaled particle theory for the porous media, Wertheim’s 
thermodynamic perturbation theory, and the Flory–Stockmayer theory. To assess the accuracy 
of the theory a set of computer simulations have been performed. 
In general, predictions 
of the theory appear to be in a good agreement with computer simulation results. 
Confinement and competition between the formation of bonds connecting the fluid
particles, and connecting fluid particles and obstacles of the matrix, 
give rise to a re-entrant phase behavior with 
three critical points and two separate regions of the liquid-gas phase coexistence.
\end{abstract}

\def\Gr{{\rm \Gamma}}
\newcommand{\be}{\begin{equation}} 
\newcommand{\ee}{\end{equation}}
\newcommand{\bea}{\begin{eqnarray}}
\newcommand{\eea}{\end{eqnarray}}

\section{Introduction}\label{sec:introduction}

Porous materials, both natural (e.g., zeolites, 
biological tissues, rocks and soil) and artificial (e.g., ceramics, synthetic zeolites, 
cement, silica aerogels), are ubiquitous. They are relevant for a broad range of 
applications, e.g., catalysis and sensing, gas storage and separation, adsorption, drug delivery, 
biomaterial immobilization, environmental remediation, molecular recognition, energy storage, etc
\cite{Corma1997,Vinu2006,Slowing2007,Ertl2008,Adiga2009,Corma2010,Stroeve2011,Misaelides2011,Dawson2012,Rouquerol2013,Liu2014,Zhang2014,Canham2014,Pal2015,Li2016,Liang2017,Kumeria2017,Danda12019,Wang2020,Miao2020,Li2021,Hosono2021,Moretta2021,Singh2021,Saha2022,Hadden2022,Yue2022,Popat2022}. In most of the earlier applications of porous materials,
conventional porous materials, such as zeolites, silica aerogels or activated carbon, were used 
\cite{Milton1989,Wang2010,Xia2019,Gu2014,Kistler1931,Kresge1992,Zhao1998}. Due to the emergence of
new classes of porous material, which include metal-organic frameworks~
\cite{Li1999, Wen2019, Easun2017, Li2009, Zhang2020}, 
porous organosilicas \cite{Mizoshita2011,Modak2010},  porous polymers \cite{McKeown2006,Wu2012,Chandra2009,Ou2018,Xu2013,Das2017} and covalent organic frameworks 
\cite{Das2017,Ding2013,Keyu2020}, substantial advances in the field were achieved during the last 
few decades. In particular, due to their well-developed porosity and tunable surface chemistry these composite 
organic-inorganic porous materials can be relatively easy functionalized, i.e. the properties of 
the inner surface of the pores can be appropriately modified by decorating it with different functional 
groups, organic and inorganic moieties, etc. Functionalized porous materials offer a number of new 
opportunities to achieve desirable properties for a diverse range of industrial applications 
\cite{Yadav2020} and thus
detailed and deep understanding of the mechanisms and effects of confinement
on the 
properties of the fluids adsorbed in the porous media is essential for their development.

Over the past several decades substantial progress in the theoretical description  of fluids adsorbed 
in disordered porous media has been achieved
 \cite{Rosinberg1999,Sarkisov2008,Holovko2013,Holovko2015PLM,Dong2018,Pizio2019}. 
Most of the theoretical studies are based on the application of the approach
developed by Madden and Glandt \cite{Madden1988,Madden1992} and Given and Stell \cite{Given1992}.
In these studies a porous medium is modeled as a quenched disordered configuration of the particles and
the properties of the fluid adsorbed in a media are described using the so-called replica Ornstein-Zernike (ROZ)
theory \cite{Madden1988,Madden1992,Given1992}. Subsequently the theory was extended and applied
to study the properties of a number of different fluids confined in the porous media, including site-site 
fluids \cite{Kovalenko2001,Chandler1991,Thompson1993}, associating fluids 
\cite{Trokhymchuk1997,Orosco1997,Pizio1998,Padilla1998,Malo1999,Urbic2004}, fluids with Coulomb interactions 
\cite{Hribar1997,Hribar1998,Hribar1999,Vlachy2004,Hribar2011}
and inhomogeneous systems \cite{Pizio1997,Kovalenko1998}. ROZ theory appears to be very useful and is able to 
provide relatively accurate predictions for the structure and thermodynamic properties of the fluid
adsorbed in disordered porous media. However, solution of ROZ equation requires application of numerical
methods, i.e. none of ROZ closures proposed so far are amenable to an analytical solution. In addition,
straightforward application of the ROZ approach for phase equilibrium calculations is limited by the
absence of a convergent solution to the ROZ equation in a relatively large region of thermodynamic states.
Recently, a theoretical approach enabling one to analytically describe the properties of hard-sphere fluid
adsorbed in the hard-sphere matrix, formed by the fluid of hard spheres quenched at equilibrium, has been 
developed \cite{Holovko2009,Chen2010,Patsahan2011,Holovko2013,Holovko2017,Kalyuzhnyi2014}. 
The theory is based on the appropriate extension of scaled particle theory~\cite{Reiss1959}.
The hard-sphere fluid is routinely used as a reference system in a number of thermodynamic perturbation theories.
In particular, a fluid of hard spheres confined in the hard-sphere matrix was used as a reference system
in the framework of the Barker-Henderson perturbation theory \cite{Holovko2015,Nelson2020}, high-temperature approximation \cite{Hvozd2017}, Wertheim's thermodynamic perturbation theory (TPT)
\cite{Kalyuzhnyi2014,Hvozd2018,Hvozd2020,Hvozd2022a,Hvozd2022b}, associative mean spherical approximation 
\cite{Holovko1991,Krienke2000,Holovko2017MSA1,Holovko2017MSA2} and collective variable method 
\cite{Holovko2016cv,Patsahan2018cv1,Patsahan2018cv2}.

In the current paper we propose a simple model for functionalized disordered porous media and study the 
effects of confinement on the clusterization, percolation and phase behavior of a fluid 
of patchy particles.
The media is represented by the disordered matrix of the hard-sphere obstacles bearing a certain number
of off-center square-well sites (patches). Due to strong, short-range attraction between patches, 
the obstacles can bond to fluid particles, as well as the fluid particles having the ability to form a network. The theoretical
description is carried out combining SPT for the porous media 
\cite{Patsahan2011,Holovko2013,Holovko2017,Kalyuzhnyi2014} and Wertheim's TPT for 
associating fluids \cite{Wertheim1986a,Wertheim1986b,Wertheim1987}. To assess the accuracy of the 
theory proposed we generate a set of computer simulation data and compare them against our 
theoretical predictions. 

The paper is organized as follows. In Section~\ref{sec:model} we describe the interaction potential model 
and present the theory. In Section~\ref{sec:simulation} we briefly describe details of the computer simulations.  
Section~\ref{sec:results} contains results and their discussion. Our conclusions are collected in Section~\ref{sec:conclusions}.

\begin{figure}[!htb]
	\centering
	\includegraphics[width=1.3cm]{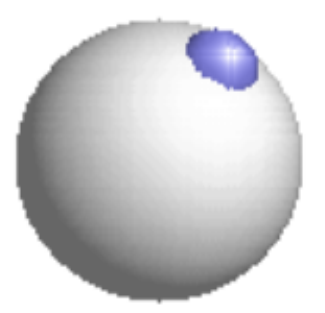} \quad
	\includegraphics[width=1.3cm]{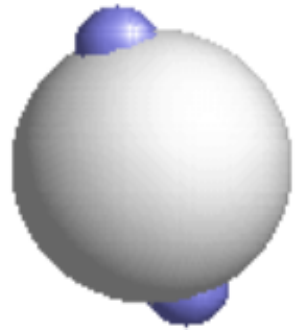} \quad
	\includegraphics[width=1.3cm]{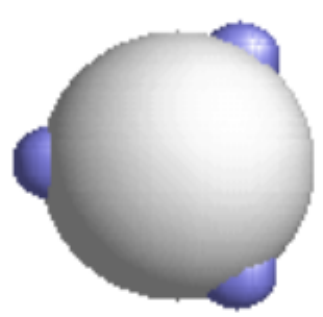} \quad
	\includegraphics[width=1.3cm]{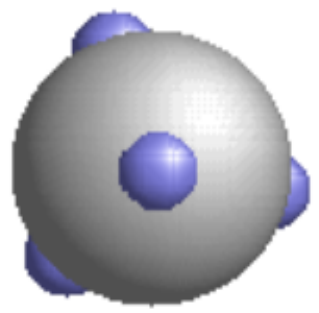}		
	\caption{Models of a particle with one, two, three and four patches.}
	\label{fig:model}
\end{figure}

\section{The Model and Theory}\label{sec:model}
 
\begin{figure}[!htb] 	
	\begin{center}
		\includegraphics[width = 4.3 cm]{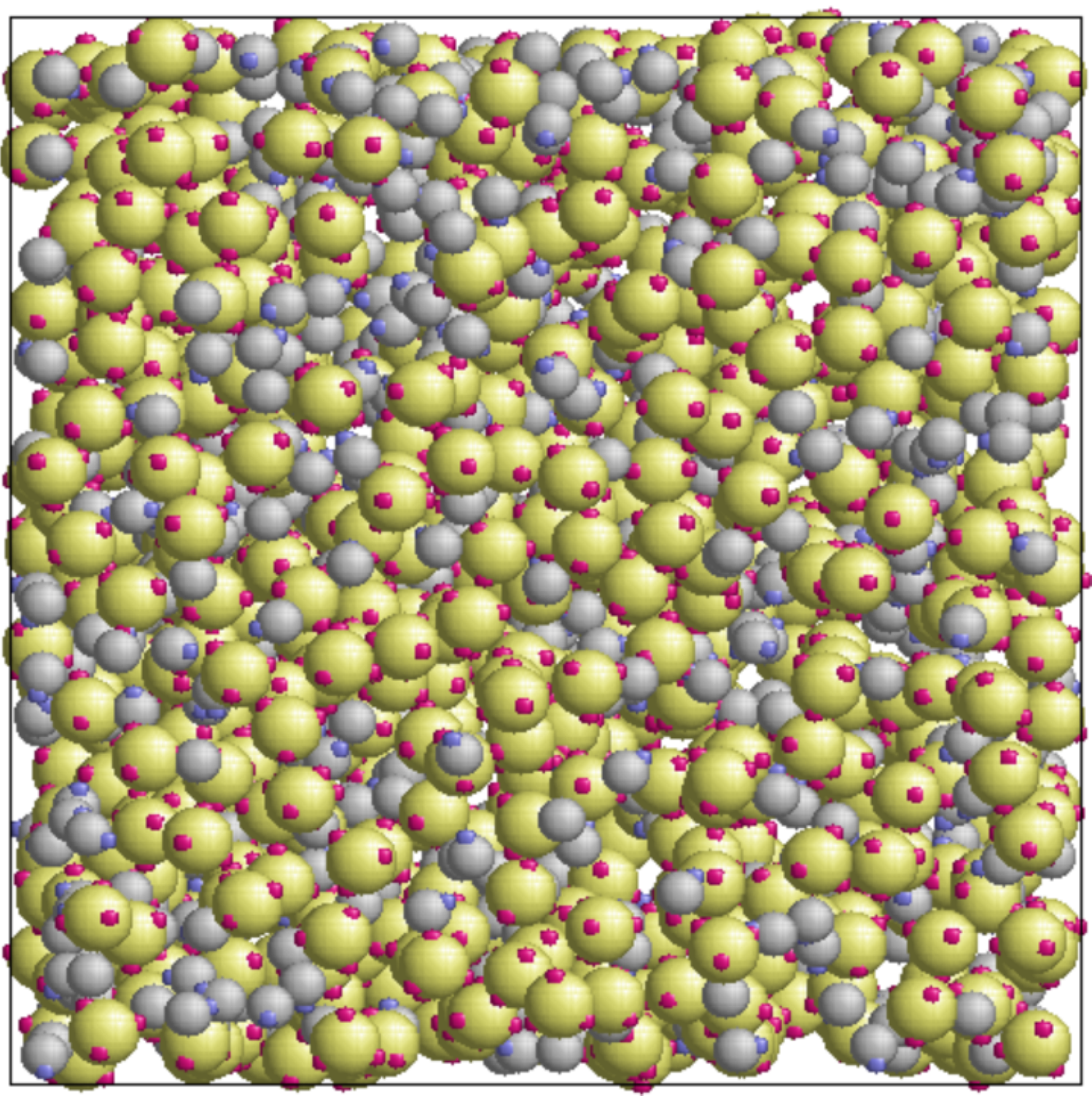} \qquad
		\includegraphics[width = 4.3 cm]{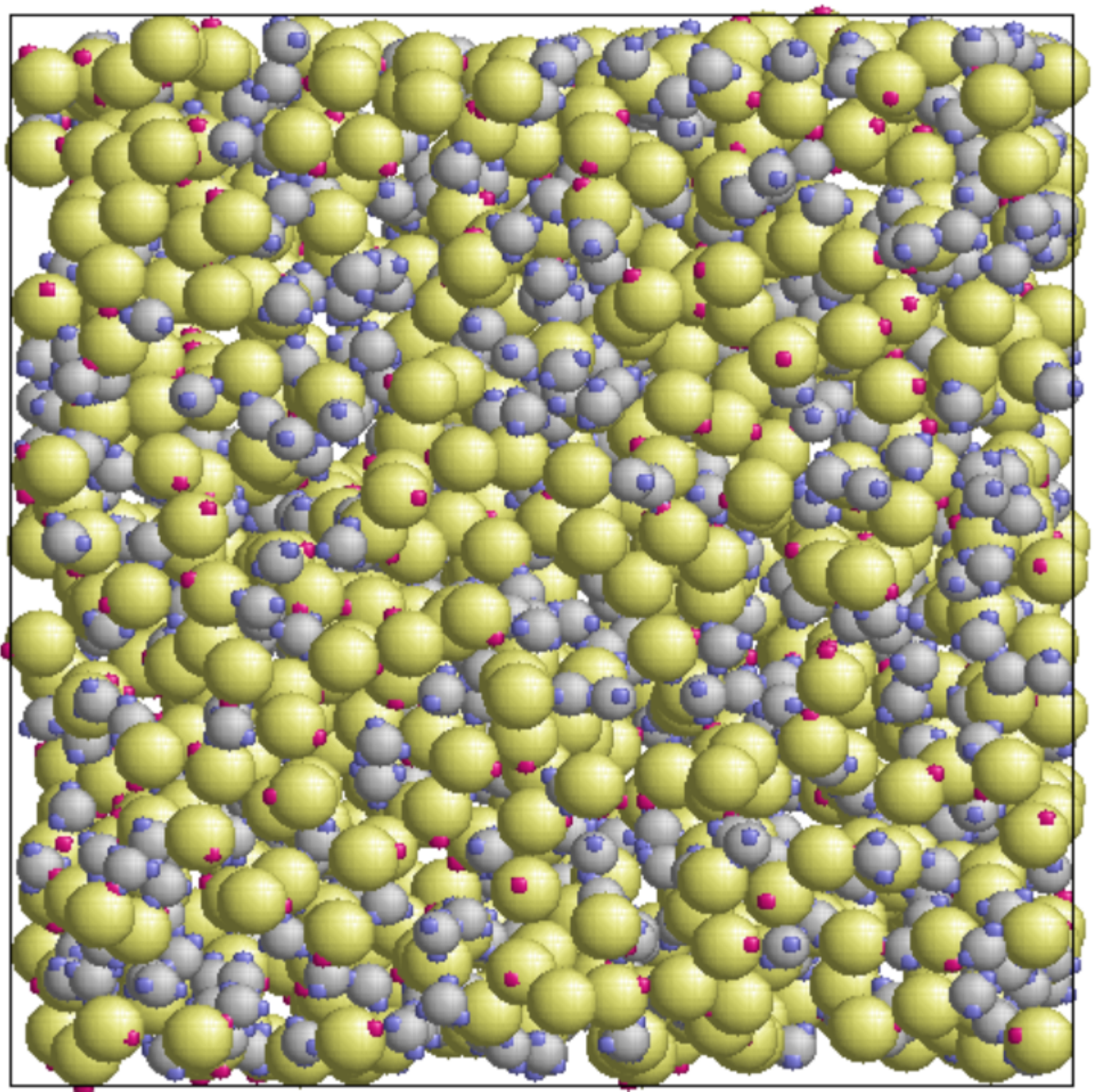}
		\caption{Models of a patchy fluid in a patchy matrix: one-patch fluid in four-patch matrix L1M4 (left panel) and 
			three-patch fluid in one-patch matrix L3M1 (right panel). {The fluid particles are represented by
				the gray spheres and matrix obstacles are shown as the green spheres.}}
		\label{fig:snaps}
	\end{center}
\end{figure}

Patchy colloids are modeled as a one-component hard-sphere fluid with the particles decorated
by the set of $n_1$ square-well bonding sites (patches) located on the surface~Fig.~\ref{fig:model}. The fluid is confined 
in the porous media represented by the matrix of the patchy hard-sphere obstacles, quenched at 
hard-sphere fluid equilibrium~Fig.~\ref{fig:snaps}. Each of the obstacles has on its surface $n_0$ square-well bonding 
sites (patches). The pair potential $U_{ij}(12)$, which describes the interaction between colloidal
particles and between colloidal particles and obstacles is given by

\be 
U_{ij}(12)=U_{ij}^{(hs)}(r)+\sum_{KL}U_{i_Kj_L}^{(as)}(12),
\label{U1j}
\ee
where 
\be
U^{(as)}_{i_Kj_L}(12)=U^{(as)}_{i_Kj_L}(z_{12})=\left\{
\begin{array}{rl}
	-\epsilon_{ij}, & {\rm for}\;z_{12}\le\omega_{ij}\\
	0, & {\rm otherwise}
\end{array}
\right.,
\label{UKL}
\ee
$1$ and $2$ stand for the position and orientation of the particles 1 and 2,
the indices $i,j$ take the values 
$(i,j)=(1,1),(1.0),(0,1)$
 and denote either fluid
particles ($i=1$) or matrix obstacles ($i=0$), the indices $K$ and $L$ denote patches and take
the values $A,B,C,\ldots$. 
$U^{(hs)}_{ij}(r)$ is the hard-sphere potential between the
particles of the sizes $\sigma_i$ and $\sigma_j$, $z_{12}$ is the distance between corresponding
sites, $\epsilon_{ij}$ and $\omega_{ij}$ are the depth and width of the square-well 
potential. Note that $\epsilon_{ij}$ and $\omega_{ij}$ are independent of the type
of the patches.

The properties of the model are calculated combining Wertheim's thermodynamic perturbation theory
(TPT) for associating fluids \cite{Wertheim1986a,Wertheim1986b,Wertheim1987} and scaled particle theory (SPT),
extended to describe a hard-sphere fluid adsorbed in the hard-sphere matrix 
\cite{Patsahan2011,Holovko2013,Holovko2017}.
According to Wertheim's TPT Helmholtz free energy of the system $A$ can be written in the 
following form
\be
A=A_{ref}+\Delta A_{as},
\label{A}
\ee
where $A_{ref}$ is Helmholtz free energy of the reference system, and
\be
\frac{\beta\Delta A_{as}}{V}=\rho_1n_1\left(\ln{X_1}-\frac{1}{2}X_1+\frac{1}{2}\right)+
\rho_0n_0\left(\ln{X_0}-\frac{1}{2}X_0+\frac{1}{2}\right),
\label{Aas}
\ee
$\rho_i$ is the number density of the particles of the type $i$, $\beta=1/(k_BT)$,
$k_B$ is the Boltzmann constant, $T$ is the temperature,
$X_i=X_{i_K}$ is fraction of the particles with attractive site of type $K$, which belong to
the particles of the type $i$, non-bonded. Note that all the patches, which belong to the particles
of the type $i$, are equivalent (\ref{UKL}). Here $X_i$ is obtained from the solution of the
following set of equations \cite{Kalyuzhnyi2014}
\cite{Wertheim1986a,Wertheim1987}
\be
4\pi X_1\left(\rho_1\sigma_{11}^3n_1X_1T_{11}^{(as)}g_{11}^{(ref)}+
\rho_0\sigma_{01}^3n_0X_0T_{01}^{(as)}g_{01}^{(ref)}\right)+X_1-1=0,
\label{X0}
\ee
and
\be
4\pi X_0\rho_1\sigma_{01}^3n_1X_1T_{01}^{(as)}g_{01}^{(ref)}+X_0-1=0,
\label{X1}
\ee
where $\sigma_{ij}=(\sigma_i+\sigma_j)/2$, 
$g_{ij}^{(ref)}$ is the contact value of the radial distribution function of the reference
system,
\be
\sigma_{ij}^3T_{ij}^{(as)}=\int_{\sigma_{ij}}^{\sigma_{ij}+\omega_{ij}}{\tilde f}_{ij}^{(as)}(r)r^2dr,
\label{TT}
\ee
and ${\tilde f}^{(as)}_{ij}(r)$ is the orientationally averaged Mayer function for square-well
site-site potential
\be
{\tilde f}^{(as)}_{ij}(r)=
\left(e^{\beta\epsilon_{ij}}-1\right)\left(\omega_{ij}+\sigma_{ij}-r\right)^2
\left(2\omega_{ij}-\sigma_{ij}+r\right)/\left(6\sigma_i\sigma_j r\right)
\label{Mayer}
\ee

\begin{figure}[!htb] 	
	\begin{center}
\includegraphics[width = 7. cm]{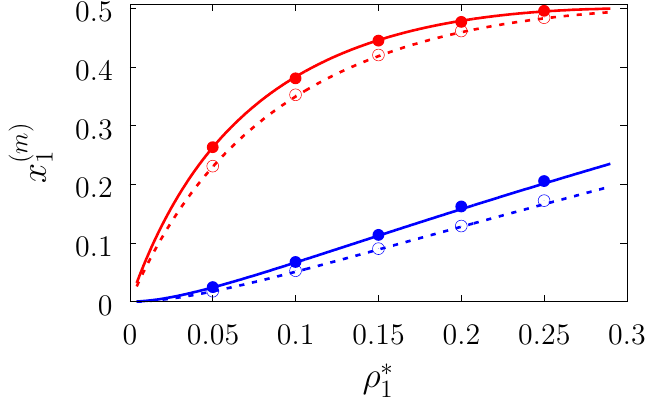}\\
\includegraphics[width = 7. cm]{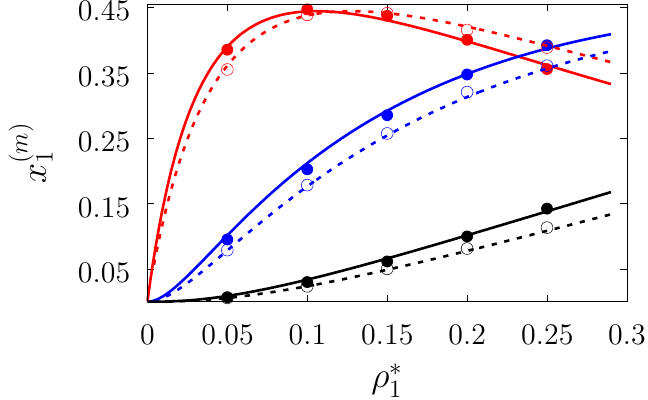}\\
\includegraphics[width = 7. cm]{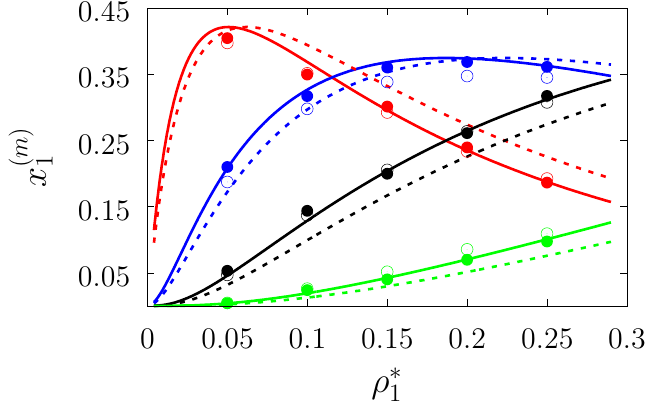}
		\caption{Fraction of $m$ times bonded fluid particles $x_{1}^{(m)}$ as a function of the fluid
			density $\rho_1^*=\rho_1\sigma_1^3$ for the model $L2M0$ (top panel), $L3M0$ (intermediate panel), 
			$L4M0$ (bottom panel) at different values of the $\eta_0$, i.e. $\eta_0=0.1$ (dashed lines
			and open circles) and $\eta_0=0.2$ (solid lines and filled circles). Here $m=1$ (red curves and symbols), $m=2$ (blue lines and symbols), 
			$m=3$ (black lines and symbols), $m=4$ (green line and symbols). Theoretical results are 
			represented by the lines and computer simulation results by the symbols.}
		\label{fig1}
	\end{center}
\end{figure}

The chemical potential, $\mu$, and pressure, $P$, of the system are calculated using standard 
thermodynamic relations, i.e.
\be
\mu=\mu_{ref}+\Delta\mu_{as},\;\;\;\;\;\;\;
P=P_{ref}+\Delta P_{as},
\label{mu}
\ee
and are used for the phase equilibrium calculations. Here
\be
\Delta\mu_{as}=
\left({\partial(\Delta A_{as}/V)\over\partial\rho_1}\right)_{T,V},
\;\;\;\;\;\;\;\;
\Delta P_{as}=\rho_1\Delta\mu_{as}-\Delta A_{as}/V.
\label{mumu}
\ee
Coexisting densities of the low-density ($\rho_1^{(l)}$) and high-density ($\rho_1^{(h)}$) phases 
at a certain temperature $T$ follow from the solution of the set of two equations:
\be
\mu(T,\rho_1^{(l)})=\mu(T,\rho_1^{(h)}),\;\;\;P(T,\rho_1^{(l)})=P(T,\rho_1^{(h)}),
\label{pec}
\ee
which represent two-phase equilibrium conditions. 

The reference system is represented by the hard-sphere fluid confined in the hard-sphere matrix.
The properties of this reference system are calculated using corresponding extension of 
scaled particle theory \cite{Patsahan2011,Holovko2013,Holovko2017,Kalyuzhnyi2014}. 
Closed form analytical expressions for Helmholtz free
energy of the reference system $A_{ref}$ and chemical potential $\mu_{ref}$ are presented in the
Appendix.  

\section{Details of computer simulations}\label{sec:simulation}
Monte Carlo computer simulations were performed for the model of patchy particles confined in a patchy hard-sphere matrix, according
to the model presented in Section~\ref{sec:model}.
We consider hard-sphere particles of fluid and matrix with diameters of $\sigma_{0}$ and $\sigma_{1}$, respectively. 
We study particles with different number of patches, which are symmetrically arranged on the surface of particle at a distance of half the diameter from its center.
As is shown in Fig.~\ref{fig:model}, the patches in a two-patch particle are located diametrically opposed to each other, in a three-patch particle they form a equilateral triangle, and in four-patch model, the patches are in vertices of a regular tetrahedron. 
Monte Carlo simulations have been carried out using the conventional Metropolis algorithm in the canonical ensemble (NVT)\cite{AllenTildesley} 
for a two-component mixture of patchy matrix and fluid particles taking into account both the translational and rotational moves of the fluid particles, 
while for the matrix particles, all degrees of freedom have been frozen.
To handle orientations and rotations of patchy particles the quaternion representation~\cite{AllenTildesley} is employed. 

The simulations have been performed in a cubic box of size $L$ with the periodic boundary condition applied.
The size of the box has been chosen to be equal to $L=24\sigma_1$, where $\sigma_1$ is a diameter of the fluid particles.
Each simulation run has been started from a random configuration of fluid particles, both in their positions and orientations. 
Configurations of matrix particles were random as well, however, they were 
the same for each of the systems studied at the given matrix density $\rho_0$ and for each set of fluid parameters.
The maximum trial displacement at each translational move of a fluid particle is limited to the distance of $0.1\sigma_1$. 
Each rotational move of a particle is performed around a random axis within the maximum rotation angle of $0.1$ radians.

\begin{figure}[!htb]	
	\begin{center}
\includegraphics[width = 7. cm]{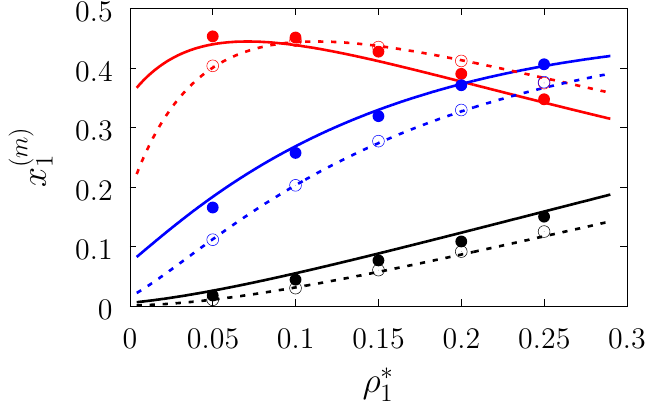}\\
\includegraphics[width = 7. cm]{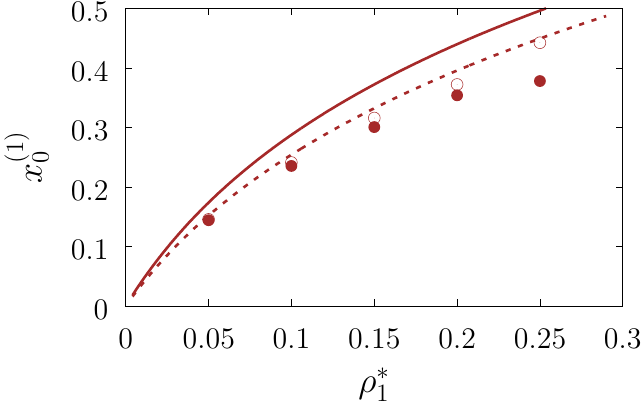}
		\caption{Fraction of $m$ times bonded fluid (top panel) and matrix (bottom panel) 
			particles $x_{i}^{(m)}$ as a function of the fluid density $\rho_1^*$ for the model $L3M1$ at different 
			values of the $\eta_0$, i.e. $\eta_0=0.1$ (dashed lines and open circles) and 
			$\eta_0=0.2$ (solid lines and filled circles). Here $m=1$ (red and brown curves and 
			symbols), $m=2$ (blue lines and symbols) and $m=3$ (black lines and symbols).
			Theoretical results are represented by the lines and computer simulation results by 
			the symbols.}
		\label{fig2}
	\end{center}
\end{figure}

The width $\omega_{ij}$ of the square-well potential is assumed to be $0.119\sigma_1$ for 
all patch-patch interactions ensuring formation of only one bond per patch.
The depth $\epsilon_{ij}$ is also taken to be the same for all the square-well 
potentials, assuming that the matrix particles are functionalized with same groups as the fluid 
particles, or at least they lead to the association between fluid and matrix particles with the 
energy close to that observed between fluid particles.
In Fig.~\ref{fig:snaps} some snapshots with examples of the simulated systems are shown.

Each simulation process was conducted in two stages: equilibration and production runs, and both runs took $10^6$ simulation steps,
where the simulation step consisted of $N$ trial translational and rotational moves of fluid particles. It appeared to be sufficient for producing reliable data,
which were obtained for two values of the packing fractions of matrix particles 
$\eta_0=\pi\rho_0\sigma_0^3/6$, i.e. $\eta_0=0.1$ and $0.2$ and different number 
densities of fluid particles ranging 
{from $\rho_1^*=\sigma_1^3N_1/V=0.05$
up to $\rho_1^*=0.25$}, where $N_{1}$ is a number 
of fluid particles and $V=L^{3}$ is a volume of the simulation box. 
The packing fraction of the matrix is an inverse characteristic to the porosity $\phi_{0}=1-\eta_0$, which is the basic quantity describing a porous medium, 
 $\rho_{0}=N_{0}/V$, $N_{0}$ is a number of matrix particles and the diameter $\sigma_{0}=1.5\sigma_{1}$ was set larger than that of the fluid particles.
In all simulations, a value of the temperature $T^{*}=k_{B}T/\epsilon_{11}$ was 
$T^{*}=0.12$, surpassing the critical temperature of the present model fluids studied in the 
bulk \cite{Bianchi2008}. This ensures that critical temperature of these model 
fluids, being confined in the matrix, is lower than $T^*=0.12$~\cite{Kalyuzhnyi2014}.
 To accelerate the simulations the linked-cells algorithm was used with a cell size of $4.0\sigma_{1}$~\cite{AllenTildesley}.
During the production run a number of bonds between fluid and matrix particles were collected at each $10$ simulation step and averaged along the simulation
in order to calculate the fraction of $m$-times bonded fluid $x_1^{(m)}$ and matrix $x_0^{(m)}$ particles and to compare them with our theoretical predictions.

\section{Results and discussion}\label{sec:results}

To systematically access accuracy of the theory we have studied the properties of several versions of the model.
We consider the models with different number of the attracting sites and different values of the
density of the fluid and matrix particles. These versions are denoted as $L_iM_j$, where $i$ and $j$ are
the number of sites on the fluid and matrix particles. We also consider two values for 
the matrix packing 
fraction and one value of the temperature, i.e. $\eta_0=0.1,\;0.2$ and $T^*=k_BT/\epsilon_{11}=0.12$, respectively.
The other parameters of the model are: $\sigma_0=1.5\sigma_{11}$, $\omega_{11}=\omega_{01}=0.119\sigma_{11}$
and $\epsilon_{01}=\epsilon_{11}$.

\begin{figure}[!htb]
	\begin{center}
		\includegraphics[width = 7 cm]{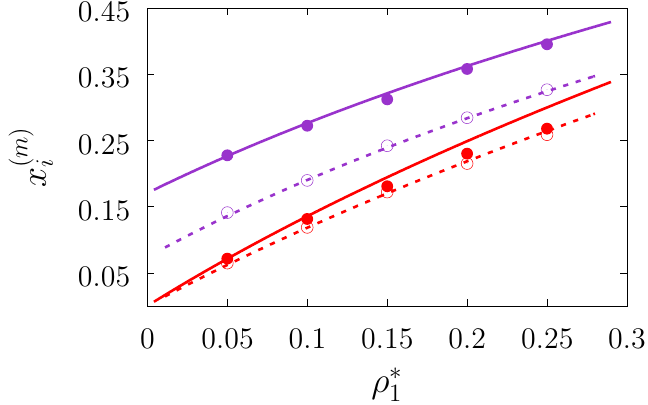}\\
		\includegraphics[width = 7 cm]{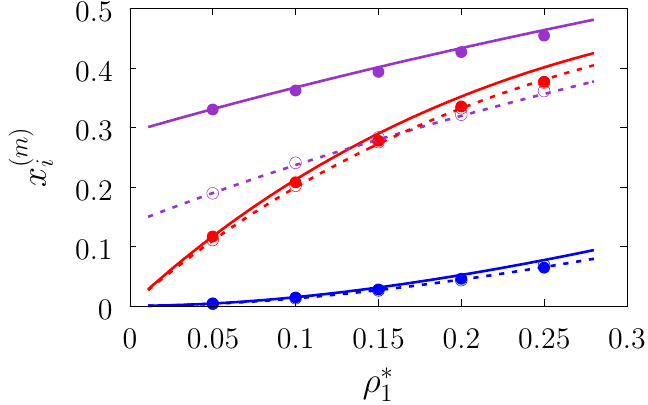}\\
		\includegraphics[width = 7 cm]{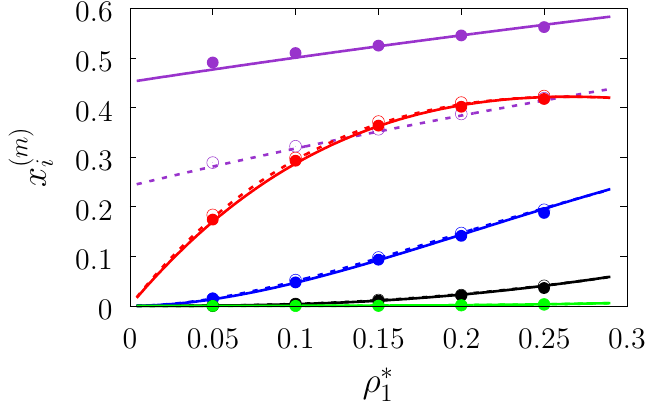}
		\caption{Fraction of $m$ times bonded fluid and matrix particles $x_{i}^{(m)}$ as a function 
			of the fluid density $\rho_1^*$ for the model $L1M1$ (top panel), $L1M2$ (intermediate 
			panel), $L1M4$ (bottom panel) at different values of the $\eta_0$, i.e. $\eta_0=0.1$ 
			(dashed lines and open circles) and $\eta_0=0.2$ (solid lines and filled circles). Here 
			fraction of singly bonded fluid particles ($m=1$) are shown by violet lines and symbols
			and for the matrix particles we have: $m=1$ (red curves and symbols), $m=2$ (blue lines 
			and symbols), $m=3$ (black lines and symbols), $m=4$ (green line and symbols). 
			Theoretical results are represented by the lines and computer simulation results by 
			the symbols.}
		\label{fig3}
	\end{center}
\end{figure}

In figures \ref{fig1}-\ref{fig3} we compare theoretical and computer simulation results for the fraction of $m$ 
times bonded fluid $x_1^{(m)}$ and matrix $x_0^{(m)}$ particles as a function of the fluid density $\rho_1$. The latter 
quantity is calculated using the following relation \cite{Wertheim1987}
\be
x^{(m)}_i=\frac{n_i^{(s)}!}{m!(n_i^{(s)}-m)!}X_i^{n_i^{(s)}-m}(1-X_i)^m,
\label{fraction}
\ee
In figure \ref{fig1} we
show our results for the models with matrix particles without patches and fluid particles with 2, 3 and 4 patches,
i.e. models $L2M0$, $L3P0$, and $L4P0$. Agreement between theoretical and computer simulation predictions for the
models $L2M0$ and $L3M0$ at both values of the matrix packing fraction, $\eta_0=0.1,\;0.2$ is very good. 
For the model $L4M0$ and lower value of $\eta_0=0.1$ results of the theory are less accurate and agreement with
exact computer simulation results is only semi-quantitative. However with the increase of the matrix packing
fraction $\eta_0=0.2$ agreement between theory and computer simulation becomes almost as good as in the case of 
the other models. 
Our results for the models with patches on both fluid and matrix particles are presented in figures \ref{fig2} 
and \ref{fig3}. In figure \ref{fig2} we compare theoretical results against computer simulation results for the
model $L3M1$. Here one can observe very good agreement between theory and simulation for $x^{(1)}_1,x^{(2)}_1$ 
and $x^{(3)}_1$ for both values of the matrix packing fraction $\eta_0$. Theoretical 
predictions for the fraction of singly bonded matrix particles $x^{(1)}_0$ are accurate for the lower 
values of $\eta_0=0.1$. At $\eta_0=0.2$ theoretical results are less accurate. While computer simulations 
predict a slight decrease in $x^{(1)}_0$ upon increase of $\eta_0$, theoretical calculations show that it increases.
{As $\eta_0$ increases the average distance between the matrix particles decreases and the number of the
obstacles with the patch blocked by the nearest neighboring obstacles increases. This effect, which is not 
accounted for in the framework of the present version of the TPT, causes slight decrease of $x_0^{(1)}$.
On the other hand this effect is less pronounced for the models $L1M1$, $L1M2$ and $L1M4$ (figure \ref{fig3}) 
and the corresponding theoretical results are in a very good 
agreement with computer simulation results}. This good agreement is observed for the fraction of $m$ times 
bonded fluid and matrix particles for both values of the matrix packing fraction $\eta_0$. 
Thus with the increase of the matrix packing fraction $\eta_0$ and number of the patches on the fluid 
particles $n_1$ the first order TPT becomes less accurate. Further improvement of the theory can
be achieved using higher-order versions of the TPT. 

\begin{figure}[!htb]
	\begin{center}
		\includegraphics[width = 8.5 cm]{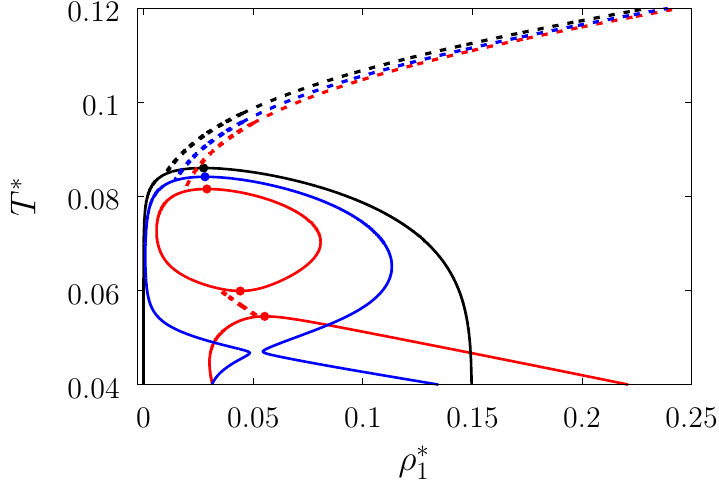}
		\caption{Liquid-gas phase diagram (solid lines) and percolation thresholds (dashed lines) for the 
			model $L1M3$ with $\omega_{01}=0.1\sigma_1$ and
			$\epsilon_{01}=0$ (black lines), $\epsilon_{01}=0.77\epsilon_{11}$ (blue lines) and 
			$\epsilon_{01}=0.825\epsilon_{11}$
			(red lines). Here symbols denote position of the corresponding critical points.}
		\label{fig4}
	\end{center}
\end{figure}

\begin{figure}[!htb]
	\begin{center}
\includegraphics[width = 7 cm]{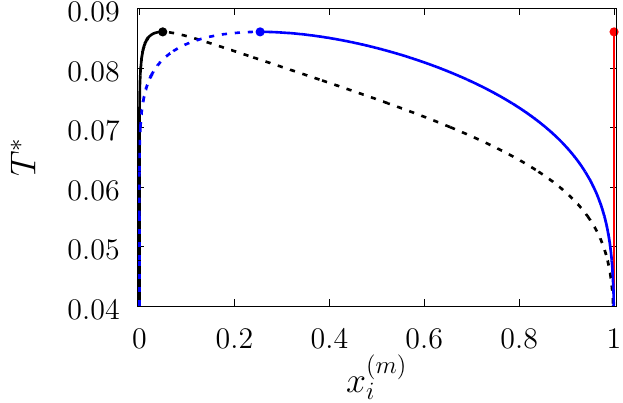}\\
\includegraphics[width = 7 cm]{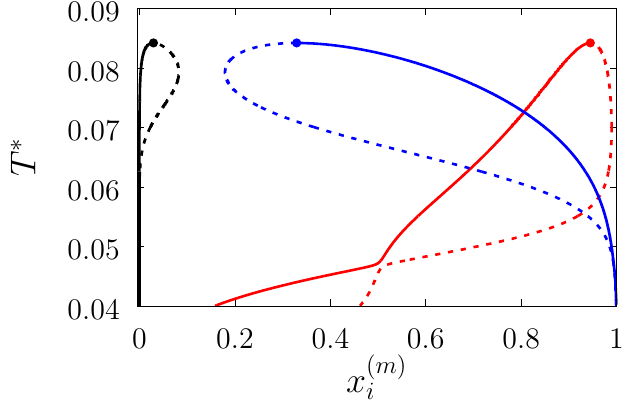}\\
\includegraphics[width = 7 cm]{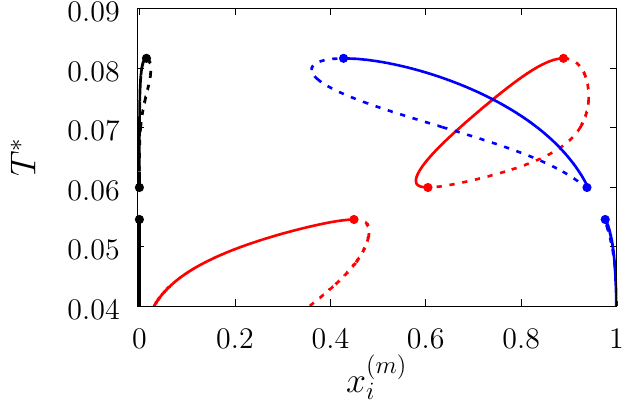}
		\caption{Fraction of $m$ time bonded fluid and matrix particles $x^{(m)}_{i}$ along coexisting lines for the 
			model $L1M3$ with $\omega_{01}=0.1\sigma_1$ and $\epsilon_{01}=0$ (top panel),  $\epsilon_{01}=0.77\epsilon_{11}$ (intermediate panel) and $\epsilon_{11}=0.825\epsilon_{11}$ (bottom panel).
			$x_1^{(0)}$ (black lines), $x_1^{(3)}$ (blue lines) and $x_0^{(0)}$ (red lines). Here dashed and solid lines 
			represent gas and liquid branches of the phase diagrams, respectively.}
		\label{fig5}
	\end{center}
\end{figure}

In figures \ref{fig4} and \ref{fig5} we present liquid-gas phase diagram and percolation threshold
lines for the model $L3M1$ with different values for the ratio between the strength of attractive 
fluid-fluid and fluid-matrix interaction. Percolation line separates $\rho\;vs\;T$ plane into percolating and
non-percolating regions, i.e. the latter region the particles form finite size clusters and in the former these
clusters form an infinite network.
Our calculation of the percolation threshold lines 
is based on the extension of the Flory-Stockmayer theory \cite{Bianchi2007,Bianchi2008,Heras2011,Tavares2010}. 
According to the earlier work \cite{Heras2011,Roldan-Vargas2013} percolation threshold lines for the model at
hand are defined by the equality $p_{11}=1/2$, where $p_{11}$ is probability of forming the bonds between the 
fluid particles, i.e.
\be
p_{11}=(1-X_1)-{\rho_0\over 3\rho_1}(1-X_0).
\label{p11}
\ee

We consider the following values 
for the hard-sphere size of the matrix obstacles, i.e. $\sigma_{01}=1.388\sigma_{11}$ and
 for the width and depth of the potential well (\ref{UKL}): $\omega_{01}=0.1\sigma_1$
and $\epsilon_{01}=0,\;0.77\epsilon_{11},\;0.825\epsilon_{11}$. All the rest of the potential
model parameters are the same as those used before. This choice of parameters allows us to study effects due to the competition between formation of the bonds connecting fluid particles and fluid particles and matrix particles.
While formation of the 3-dimensional network of bonds connecting the fluid particles causes gas-liquid phase separation,
bonding of the fluid and matrix particles suppress it. For the model with $\epsilon_{01}=0$ the phase diagram is of the
usual form, i.e. the difference between the coexisting densities of the gas and liquid phases increases on
decreasing temperature (figure \ref{fig4}).
Here, since the density of the gas phase $\rho_{1,gas}$ is lower than the density of the liquid phase $\rho_{1,liq}$ 
the fraction of the free (or three time bonded) particles in the gas phase $x_{1,gas}^{(0)}$ ($x_{1,gas}^{(3)}$) 
is larger (lower) than corresponding fractions in the liquid phase (figure \ref{fig5}, top panel). 
As one would expect for the fractions of free matrix particles we have
$x_{0,gas}^{(0)}=x_{0,liq}^{(0)}=1$, 
where $x_{0,gas}^{(0)}$  and $x_{0,liq}^{(0)}$ denote fractions of free (nonbonded)
particles of the matrix, which confine fluid particles either in the gas phase or in the liquid 
phase, respectively.
Increase of $\epsilon_{01}$ causes changes of the phase diagram topology.
For the model with $\epsilon_{01}=0.77\epsilon_{11}$ the phase diagram exhibits 
re-entrant phase behavior, i.e. decrease of the temperature in the range of approximately $0.065 \geq T^* \geq 0.046$
results in the reduction of the difference in coexisting gas and liquid densities (figure \ref{fig4}). This effect is caused by the increased bonding of the fluid particles and matrix obstacles, 
which destroys the three-dimensional network formed by the fluid particles and suppresses the phase separation.
In this range of temperatures the fraction of free obstacles $x_0^{(0)}$ in both phases rapidly decreases (figure \ref{fig5}, intermediate panel). 

At the same time, while the fractions of free and three times bonded
fluid particles in the liquid phase do not change much, in the gas phase they are approaching the liquid phase
values (figure \ref{fig5}, intermediate panel). This behavior reflect the process of breaking the bonds between the fluid
particles and their substitution by the bonds formed between the fluid particles and obstacles of the matrix.
Upon reaching sufficient bonding degree of the matrix obstacles this process slows down and with further decrease of the 
temperature ($T^*<0.046$), the three-dimensional network formed by the fluid particles becomes stronger. As a result
the difference between coexisting gas and liquid densities increases with the temperature decrease (figure \ref{fig4}).
Similar behavior can be observed for the model with $\epsilon_{01}=1.835\epsilon_{11}$. However stronger fluid-matrix
attraction completely suppresses the phase separation in the range of the temperatures $0.060 \leq T^* \leq 0.0545$.
As a result we have the phase diagram with two separate regions of the liquid-gas phase coexistence: the region 
at higher temperatures has the upper and lower critical points (figures \ref{fig4} and \ref{fig5}) and the region
at lower temperatures has the usual liquid-gas type of the critical point. 
These critical points and the liquid branches of the phase diagrams are located inside percolating region 
(figure \ref{fig4}). The corresponding percolation line consists of two branches, i.e. high-temperature branch and 
low-temperature branch. The former branch is located at higher temperatures with cluster fluid phase above and
percolating fluid phase below percolation threshold line. The latter branch is located at lower temperatures 
and connects high-temperature and low-temperature regions of the phase diagram. Here the percolating fluid phase
occurs above and cluster fluid phase below percolation threshold line. For the models with 
$\epsilon_{01}=0\;\rm{and}\;0.77\epsilon_{11}$ percolation threshold lines are of the usual shape and almost 
completely coincide with the high-temperature percolation threshold line for the model with 
$\epsilon_{01}=0.825\epsilon_{11}$.
Finally, we note that recently the phase behavior of the patchy colloids adsorbed in the attractive
random porous medium was studied \cite{Hvozd2022a,Hvozd2022b}. In these studies the medium was represented as a 
matrix of hard-sphere obstacles with attractive Yukawa interaction between the centers of the fluid and matrix 
particles. Similar to the present study, confinement causes the system to undergo re-entrant liquid-gas phase 
separation. However competition between Yukawa interaction and bonding do not cause the phase diagram to be 
separated into two (or more) regions.
Taking into account the good agreement between the theoretical and computer simulation predictions for 
the fractions $X_0$ and $X_1$ and the fact that these are the only quantities, that enter into expressions for 
Helmholtz free energy $\Delta A_{as}$ (\ref{Aas}) and probability of forming the bonds $p_{11}$ (\ref{p11}) 
there are good reasons to believe that our theory is accurate enough to at least qualitatively 
predict the peculiarities of the phase behavior discussed above. In a subsequent paper we are planning to 
systematically investigate the phase behavior and percolation properties of the model using an
extension of the aggregation-volume-bias MC method \cite{chen2001improving,russo2011reentrant}
and the theory developed herein.

\section{Conclusions}\label{sec:conclusions}

In this work we propose a simple model of functionalized disordered porous media represented by the matrix
of hard-sphere fluid particles quenched at equilibrium and decorated by the certain number of the off-center 
square-well sites. The model is used to study effects of confinement on the clusterization, percolation and 
phase behavior of the fluid of patchy particles adsorbed in such porous media. The study is
carried out combining Wertheim's TPT for associating fluids, SPT for the porous media and Flory-Stockmayer
theory of polymerization. A set of computer simulation data has been generated and used to assess the 
accuracy of the theory. Very good agreement between theoretical and computer simulation predictions 
is observed for the fractions of $m$-times bonded fluid and matrix particles almost in all cases studied.
Slightly less accurate are results for the fraction of $m$-times bonded matrix particles for the model with
fluid particles bearing more than one patch.
Further improvement of the theory can be achieved using higher-order versions of the TPT.

The liquid-gas phase diagram and percolation threshold line for the model with three-patch fluid particles and 
one-patch matrix particles are calculated and analyzed. It is demonstrated that confinement can substantially
change the shape of the phase diagram. Bonding between the fluid particles causes formation of a network
and bonding of the fluid particles to the obstacles suppresses this process. Competition between these two
effects define the shape of the phase diagram and gives rise to a re-entrant phase behavior with three 
critical points and two separate regions of the liquid-gas phase coexistence.

\section*{Acknowledgements}
YVK, TP and MH gratefully acknowledge financial support of the National Research Foundation of 
Ukraine (project No.2020.02/0317). One of us (YVK) would like to express his gratitude
to the Vanderbilt University, where part of this research was conducted, for hospitality.



\renewcommand\refname{References}

\bibliography{refs} 
\bibliographystyle{model6-num-names}


\newpage
\section*{Appendix}
\setcounter{equation}{0}
\renewcommand{\theequation}{A\arabic{equation}}
\subsection*{\label{App:A}Expressions for Helmholtz free energy, chemical potential and pressure of the hard-sphere fluid confined in the hard-sphere matrix}

The reference system is represented by the hard-sphere fluid confined in the hard-sphere matrix.
{According to SPT~\cite{Patsahan2011,Holovko2013,Holovko2017,Kalyuzhnyi2014} we have
the following expressions for Helmholtz free energy $A_{hs}$ and for the contact values of
the radial distribution functions $g^{hs)}_{10}(r)$ and $g^{(hs)}_{11}(r)$ for hard-sphere 
fluid adsorbed in the hard-sphere matrix, i.e.}
 
\begin{equation}
\beta\Delta A_{hs}\over N_{hs}=\beta\Delta\mu_{hs}-{\beta \Delta P_{hs}\over\rho_{hs}}
\label{Ahs}
\end{equation}
and
\begin{equation}
g_{1j}^{(hs)}={1\over \phi_0-\eta_1}+{3\over 1+\tau_{1j}}
	{(\eta_1+\eta_0\tau_{10})\over (\phi_0-\eta_1)^2}
\label{g}
\end{equation}

where $N_{hs}$ is the number of hard-sphere particles of the fluid, 
$$
{\beta\Delta P_{hs}\over\rho_{hs}}={1\over 1-\eta_1/\phi_0}{\phi_0\over\phi}
+\left({\phi_0\over\phi}-1\right){\phi_0\over\eta_1}\ln\left(1-{\eta_1\over\phi_0}\right)
$$
\begin{equation}
+{a\over 2}{\eta_1/\phi_0\over(1-\eta_1/\phi_0)^2}+
{2b\over 3}{(\eta_1/\phi_0)^2\over(1-\eta_1/\phi_0)^3}-1,
\label{P}
\end{equation}
$$
\beta\Delta\mu_{hs}=
\beta\mu^{(ex)}_{1}
-\ln\left(1-{\eta_1\over\phi_0}\right)
+
{\eta_1(\phi_0-\phi)\over\phi_0\phi(1-\eta_1/\phi_0)}
+\left(1+a\right){\eta_1/\phi_0\over(1-\eta_1/\phi_0)}
$$
\begin{equation}
+{(a+2b)\over 2}{(\eta_1/\phi_0)^2\over(1-\eta_1/\phi_0)^2}
+{2b\over 3}{(\eta_1/\phi_0)^3\over(1-\eta_1/\phi_0)^3},
\label{mu}
\end{equation}
$\eta_0=\pi\rho_0\sigma_0^3/6$, $\phi_0=1-\eta_0$, $\eta_1=\pi\rho_{1}\sigma_{1}^3/6$
and $\phi=\exp{(-\beta\mu^{(ex)}_{1})}$.

\noindent
Here
\begin{equation}
a=6+{3\eta_0\tau_{10}\left(\tau_{10}+4\right)\over 1-\eta_0}+
{9\eta_0^2\tau_{10}^2\over(1-\eta_0)^2},
\;\;\;\;\;\;\;\;\;\;\;\;
b={9\over 2}\left(1+{\tau_{10}\eta_0\over 1-\eta_0}\right)^2,
\label{b1}
\end{equation}
$$
\beta\mu_{1}^{(ex)}=-\ln{(1-\eta_0)}+{9\eta_0^2\over 2(1-\eta_0)^2}-\eta_0Z_0
+\left[3\eta_0Z_0-{3\eta_0(2+\eta_0)\over(1-\eta_0)^2}\right](1+\tau_{10})
$$
\begin{equation}
-\left[3\eta_0Z_0-{3\eta_0(2+\eta_0)\over 2(1-\eta_0)^2}\right](1+\tau_{10})^2
+\eta_0Z_0(1+\tau_{10})^3,
\label{mu1}
\end{equation}
$Z_0=(1+\eta_0+\eta_0^2)/(1-\eta_0)^3$, 
{
$\tau_{1j}=\sigma_1/\sigma_j$ and $j=0,1$.
}

\end{document}